\documentclass[a4paper]{jpconf}
\usepackage{amsfonts}
\usepackage{amsmath}
\usepackage{epsfig}
\usepackage{float}
\usepackage{enumerate}
\usepackage{lmodern}

\newcommand{\RR}{{\mathbb R}}
\newtheorem{defn}{Definition}

\begin{document}

\title{Quantum mechanics from invariance principles}

\author{Florin Moldoveanu}

\address{Logic and Philosophy of Science Research Group, University of Maryland, College Park, MD 20742, USA}
\ead{fmoldove@gmail.com}

\begin{abstract}
Quantum mechanics is an extremely successful theory of nature and yet it lacks an intuitive axiomatization. In contrast, the special theory of relativity is well understood and is rooted into natural or experimentally justified postulates. Here we introduce an axiomatization approach to quantum mechanics which is very similar to special theory of relativity derivation. The core idea is that a composed system obeys the same laws of nature as its components. This leads to a Jordan-Lie algebraic formulation of quantum mechanics. The starting assumptions are minimal: the laws of nature are invariant under time evolution, the laws of nature are invariant under tensor composition, the laws of nature are relational, together with the ability to define a physical state (positivity). Quantum mechanics is singled out by a fifth experimentally justified postulate: nature violates Bell's inequalities.
\end{abstract}

Quantum mechanics is a very unintuitive theory: it predicts only probabilistic outcomes but it supposes to be the ``whole story'' \cite{EPR1}, it exhibits correlations between separable systems which cannot be explained by ``local realistic'' means \cite{Bell1}, it is based on an abstract formalism involving hermitean operators and complex vector spaces. Also it is not even clear what we mean by ``quantum'' \cite{PR1}. 

In contrast, the special theory of relativity is well understood and is rooted in two simple postulates: the laws of nature are invariant under changes in inertial frames of reference and the principle of invariant light speed. The first postulate of the special theory of relativity is easy to understand. The second postulate is ultimately justified by experimental evidence (starting with the Michelson - Morley experiment \cite{MMExperiment}). 
   
It is the aim of this paper to start deriving quantum mechanics in a very similar way to how the special theory of relativity is obtained. 

\begin{figure}[ht]
\centering
\epsfig{file=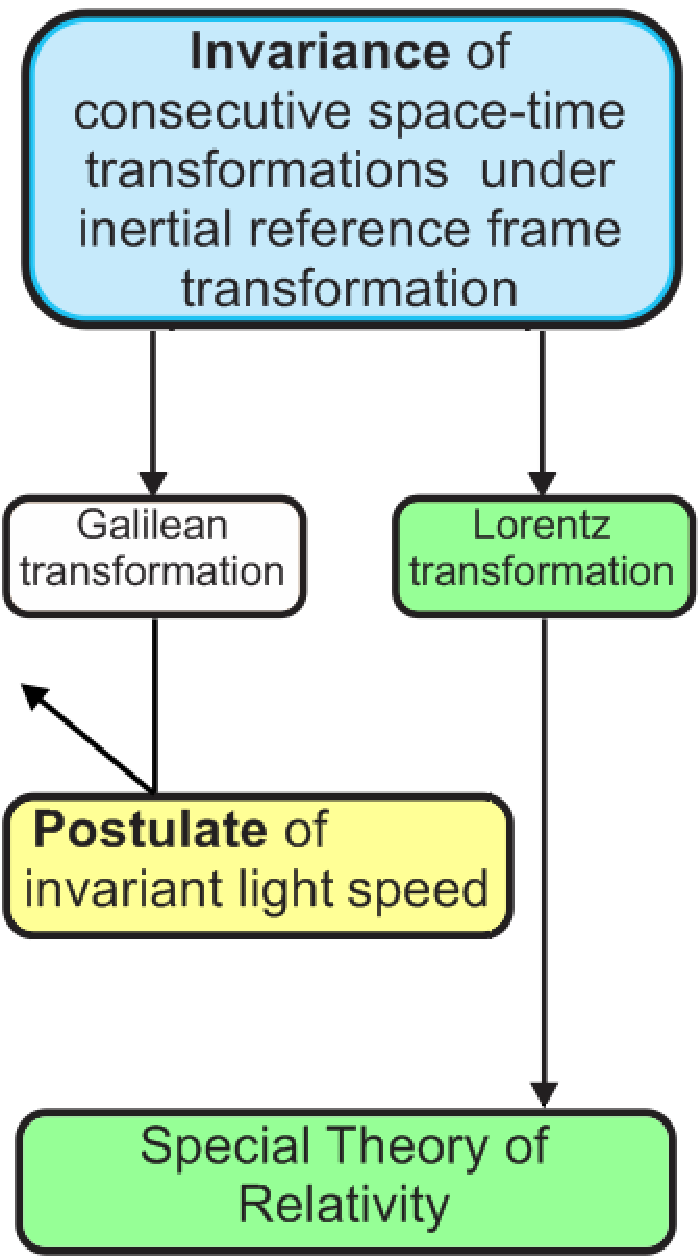,width=0.2\linewidth,clip=}
\caption{Deriving the special theory of relativity line of argument.}
\label{fig:LineOfArgumentSTR}
\end{figure}

The special theory of relativity has a kinematical foundation, but emphasizing this fact obscures a larger point that it is based on a specific invariance of the laws of nature. In particular, the special theory of relativity uses only one kind of invariance, related of inertial reference frames. Figure~\ref{fig:LineOfArgumentSTR} presents one possible line of argument for deriving the special theory of relativity.

\begin{figure}[ht]
\centering
\epsfig{file=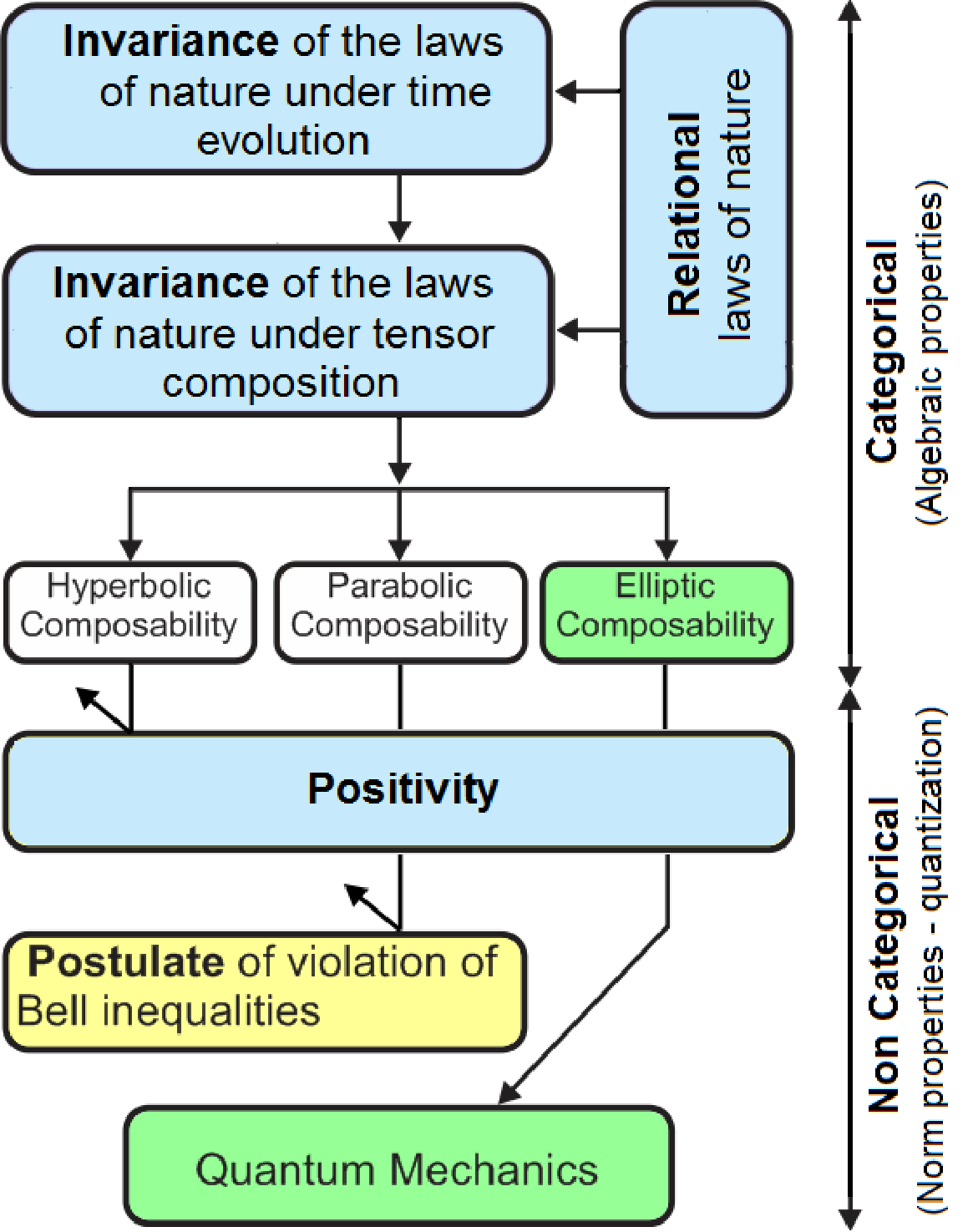,width=0.3\linewidth,clip=}
\caption{Deriving quantum mechanics line of argument.}
\label{fig:LineOfArgumentQM}
\end{figure}

Similarly, quantum mechanics can be derived from several other invariances and from natural or experimentally justified postulates (see Figure~\ref{fig:LineOfArgumentQM}).

When talking about dynamics, we need to allow interaction between any two physical systems. The critical element is the invariance of dynamics under tensor composition of two subsystems. Composability was originally discoverd in the 1970s by Emile Grgin and Aage Petersen \cite{GrginCompPaper} as the necessary ingredient of understanding classical and quantum mechanics in a single mathematical formalism. The original motivation was a belief by Bohr (as reported by his personal assistant Aage Petersen) that the correspondence principle has more to reveal. This grounbreaking work resulted in what is now called a Jordan-Lie algebra \cite{LandsmanBook}. 

The Jordan-Lie algebra requires a norm and an important observation is that this norm is unique \cite{ConnesBook} given the spectral distance. Therefore to derive quantum mechanics we have to derive the necessity of the Jordan algebra of observables and the Lie algebra of generators together with their compatibility condition. From the algebraic properties of the Jordan-Lie structure and positivity, the usual Hilbert space formulation is recovered in special cases using Berezin deformation quantization \cite{BerezinQuantization}. 

In a Jordan-Lie approach to quantum mechanics, the starting point is the existence of two products, one describing the generators and the other one the observables, which form a duality known as ``the equivalence of observables and generators'' \cite{GrginDualityPaper} or ``dynamic correspondence'' \cite{AlfsenShultz}. However, demanding the two products is too strong a requirement and we can start from much milder assumptions arriving at them. 

The starting point is requiring the existence of time and a configuration space manifold $Q$ endowed with a product (a tensor product) with unit. 

At a point $p\in Q$ one can define a tangent space $T_p Q$ and a cotangent space ${T_p}^{*} Q$. We will also assume that there are some $C^{\infty}$ functions over the cotangent bundle $M$ for which there is a way to generate a vector field out of them, and from now on we will restrict the domain of discussion only for those functions. 

Let time evolution be represented by a one parameter group of transformations $\phi$ defined as follows:

\begin{equation}
\phi : M \times \mathbb{R} \rightarrow M ,
\end{equation}

\noindent with $\phi(x, 0) = x$ and $\phi(\phi(x,t),s)=\phi(x, t+s)$.

Suppose that there is an unspecified family of local operations $\{ \circ \}$ which describe the laws of nature (for example Poisson bracket, Jordan product, commutator, etc). Introducing the notation: $\phi_t (x) \equiv \phi(x,t)$, the invariance of the laws of nature under time evolution reads:

\begin{equation}
(g \circ h) (\phi_{\Delta t}(p)) =  g (\phi_{\Delta t}(p)) \circ h (\phi_{\Delta t}(p)) ,
\end{equation}

\noindent with $p \in M$ a point in the cotangent bundle $M$ and $g,h$, functions defined in the neighborhood of $p$. In other words, we demand the existence of a universal local morphism which preserves all algebraic relationships under time translation.

If $X^i_f$ is a vector field arising out of a function $f$ (corresponding to a particular time evolution), define $\mathcal{T}_{f_\epsilon}$:

\begin{equation}
\mathcal{T}_{f_\epsilon} = I + \epsilon X_f^i \frac{\partial}{\partial u^i} ,
\end{equation}

\noindent with $I$ the identity operator and  $u^i$ the coordinate set in a local $\mathbb{R}^{2n}$ chart covering the point $p\in M$. 

Because $f$ is in one-to-one correspondence with $X_f^i$ we can introduce a time translation transformation $T_{f_\epsilon}$ and a product $\alpha \in \{ \circ \} $ between a distinguished $f$ and any $g$ as follows:

\begin{equation}
f \alpha g = \lim_{\epsilon \rightarrow 0} \frac{g - T_{f_\epsilon} g}{\epsilon} ,
\end{equation}

\noindent which is the Lie derivative of $g$ along the vector field generated by $f$ corresponding to a particular time evolution. Equivalently $T_{f_\epsilon} = (I - \epsilon f \alpha \cdot)$.

We generalize the product $\alpha$ for all $f$'s and $g$'s by repeating the argument for all conceivable dynamics. To make sure the domains of $f$ and $g$ are identical and well behaved, in case of pathologies, we can restrict the set of $g$ to the span of all possible $f$.

Then the invariance of the laws of nature under time evolution can be expressed as:

\begin{equation}
T_{f_\epsilon} (g \circ h) = [T_{f_\epsilon} g] \circ [T_{f_\epsilon} h] ,
\end{equation}

\noindent which to first order in $\epsilon$ implies a left Leibniz identity:

\begin{equation}
f \alpha (g \circ h) = (f \alpha g) \circ h + g \circ (f \alpha h) .
\end{equation}

From the relational property of the laws of nature, we demand for any $f$ that  $1\alpha f = 0$. Also we have $f\alpha 1 = 0$ from the definition of the Lie derivative. $\alpha$ will be shown later to be the usual commutator in quantum mechanics (or the Poisson bracket in classical mechanics) but it has no skew-symmetry property yet.

When talking about dynamics, it is usual to consider the interaction of two systems. The key property of a configuration space of many systems is the ability to concatenate the sub-systems using a tensor product. 

Inspired by Grgin's groundbreaking work \cite{GrginCompPaper} we can introduce a composability category $\mathcal{U}=\mathcal{U}(\otimes, \mathbb{R}, \alpha, \cdots )$ which respects the existence of a unit element, the real numbers field $\mathbb{R}$ (understood as the set of constant functions), meaning $\mathcal{U}\otimes\mathbb{R} \simeq \mathcal{U} \simeq \mathbb{R} \otimes \mathcal{U}$. 

First it can be shown that the product $\alpha$ is not enough and this demands the existence of a secondary product $\sigma \in \{ \circ \}$. The invariance of the laws of nature under composability demands that a bipartite product $\alpha_{12}$ should be built out of the products listed in the composability category. If only a product $\alpha$ exists in the composability category, it must be trivial due to Leibniz identity. Only by adding another product $\sigma$ we can construct non-trivial mathematical structures.

The most general way to construct the products $\alpha$ and $\sigma$ in a bipartite system is as follows:

\begin{eqnarray}
(f_1 \otimes f_2) \alpha_{12} (g_1 \otimes g_2) = a (f_1 \alpha g_1) \otimes (f_2 \alpha g_2) + \\ \nonumber
b (f_1 \alpha g_1) \otimes (f_2 \sigma g_2) + c (f_1 \sigma g_1) \otimes (f_2 \alpha g_2) + \\ \nonumber
d (f_1 \sigma g_1) \otimes (f_2 \sigma g_2)  
\end{eqnarray}

\begin{eqnarray}
(f_1 \otimes f_2) \sigma_{12} (g_1 \otimes g_2) = x (f_1 \alpha g_1) \otimes (f_2 \alpha g_2) + \\ \nonumber
y (f_1 \alpha g_1) \otimes (f_2 \sigma g_2) + z (f_1 \sigma g_1) \otimes (f_2 \alpha g_2) + \\ \nonumber
w (f_1 \sigma g_1) \otimes (f_2 \sigma g_2)  
\end{eqnarray}

\noindent with $f_1 , f_2 , g_1 , g_2$ arbitrary functions over the manifold $M$ at a point $p$.

Then we can use the existence of the unit of the composability category to determine the coefficients $b,c,d,y,z,w$.  If we normalize the definition of product $\sigma$ such that $1 \sigma 1 = 1$, we can show that $d=y=z=0$ and $b=c=w=1$. Applying the Leibniz identity on the bipartite products demands $a=0$. Hence the bipartite products in the shorthand notation are: 

\begin{eqnarray}\label{FundamentalRelation}
\alpha_{12}   &=& \alpha_1 \sigma_2 + \sigma_1 \alpha_2 \\ \nonumber
\sigma_{12} &=& \sigma_1 \sigma_2 + x \alpha_1 \alpha_2
\end{eqnarray}

We can now observe that if $\alpha$ is a skew-symmetric product and $\sigma$ is a symmetric product the symmetry and skew-symmetry is maintained under composition. 

So far the product $\alpha$ is undefined. We know it is a derivation because it obeys the Leibniz identity, meaning it forms a Loday algebra \cite{LodayAlg}, but it can be further proved that the product $\alpha$ is skew-symmetric. In turn this shows that the product $\sigma$ is symmetric.

Let us summarize what we have up to this point. First, there is a skew-symmetric bilinear product $\alpha$ which obeys both a Leibniz identity and a Jacobi identity. As such it forms a Lie algebra. Then there is a symmetric bilinear product $\sigma$ and a universal parameter $x$ which is a constant of nature. The normalized parameter $x$ can be $-1$, $0$, or $+1$ corresponding to elliptic, parabolic, or hyperbolic composability classes. In quantum mechanics case $x=-\hbar^2/4$ and the fact that the Planck constant is invariant is a non-trivial fact due to composability \cite{SahooPlank}. A side-effect of the invariance of the Planck constant as a consequence of composability is the impossibility to have a consistent theory of mixed classical-quantum world because classical and quantum mechanics belong in disjoint composability classes \cite{SahooPlank}. 

From $\sigma_{12} = \sigma_1 \sigma_2 + x \alpha_1 \alpha_2$ it is easy to see that there is a linear transformation $J$ from the space of observables  to the space of generators called `dynamic correspondence'' \cite{AlfsenShultz} or ``the equivalence of observables and generators'' \cite{GrginDualityPaper} such that $J \sigma J = \sqrt{x} I$ . 

This map is the root cause of the existence of complex numbers in all quantum mechanics formulations even when quantum mechanics is represented over real numbers \cite{AdlerQuaternions}. 

We can now investigate the relationship between the product $\alpha$ and the product $\sigma$.

In general the products $\alpha$ and $\sigma$ are not necessarily associative. For an arbitrary product $*$, a measure of non-associativity is the associator:
\begin{equation}
[f,g,h]_{*} = (f*g)*h - f*(g*h)
\end{equation}
Using the Jordan and Leibniz identities along with the skew-symmetry of the product $\alpha$, one can show \cite{GrginCompPaper} that there is a relationship between the $\alpha$ associator and $\sigma$ associator, called the Petersen identity:
\begin{equation}\label{PetersenIdentity}
[f,g,h]_{\sigma} + J [f,g,h]_{\alpha} = 0
\end{equation}

In turn this means that an associative product $\beta$ can be introduced as: $\beta = \sigma + J \alpha$.
 
When $J \neq 0$, in the quantum case, by choosing $f=h$ and $f=h\sigma h$ in Eq.~(\ref{PetersenIdentity}) and using the Leibniz and skew-symmetry property of the product $\alpha$ we obtain: $[h,g,h]_{\sigma} = 0$ and $[h\sigma h, g,h]_{\sigma} = 0$. This means that the product $\sigma$ obeys the flexible law and Jordan identity \cite{GrginCompPaper}. 

For the classical case, the product $\sigma$ is always an associative product in addition to being commutative (symmetric).

A direct consequence of those results implies that quantum mechanics cannot be formulated over octonions because the product $\beta$ has to be associative and therefore the Jordan algebra of observables $\sigma$ cannot be special. The classification of real Jordan algebras restricts the allowed number systems for quantum mechanics.

We can also confirm that the tensor product is associative by verifying that: $\alpha_{(12)3} = \alpha_{1(23)}$ and $\sigma_{(12)3} = \sigma_{1(23)}$ by direct application of Eq.~(\ref{FundamentalRelation}). Similarly we can see that the tensor product is also commutative: $\alpha_{(12)} = \alpha_{(21)}$ and $\sigma_{(12)} = \sigma_{(21)}$, thus making the composability category a commutative monoid. 

If in quantum mechanics one changes the imaginary unit of complex numbers from $\sqrt{-1}$ to $\sqrt{+1} \neq \pm 1$, one obtains the so called ``hyperbolic quantum mechanics'' over split-complex numbers \cite{HyperbolicQMPaper}. This corresponds to the hyperbolic composability case, and this case violates the Stone-von Neumann theorem because it has non-equivalent representations even in the finite dimensional case. 

More important, it can be proved that in the hyperbolic case one encounters negative probabilities and the hyperbolic case can be eliminated by requiring positivity. 

Therefore the only physical cases remaining are classical mechanics with the product $\alpha$ the Poisson bracket, and the product $\sigma$ the point-wise function multiplication, and quantum mechanics with the product $\alpha$ the commutator and the product $\sigma$ the anti-commutator. By composability, nature can only be in one composability class, and the way in which we can distinguish in nature between the elliptic and parabolic composability classes is by experimental evidence.    

It is well known that classical mechanics obeys Bell inequalities \cite{Bell1} and quantum mechanics can achieve higher correlations. The strong experimental evidence in favor of violations of Bell inequalities starts with the Aspect experiment \cite{AspectExperiment}. 

Collecting all the results above we introduce the following definition:

{\defn A composability two-product algebra is a real vector space $\mathfrak{A}_{\RR}$ equipped with two bilinear maps $\sigma$ and $\alpha$ such that the following conditions apply:
\begin{eqnarray*}
\alpha {\rm ~is~a~Lie~algebra} ,\\
\sigma {\rm ~is~a~Jordan~algebra} ,\\
\alpha {\rm ~is~a~derivation~for~}\sigma {\rm ~and~} \alpha ,\\
{[ A, B, C] }_{\sigma} + \frac{J^2 \hbar^2}{4} {[A, B, C]}_{\alpha} = 0 ,
\end{eqnarray*}
where $J \rightarrow (-J)$ is an involution, $1\alpha A = A\alpha 1 = 0$, $1\sigma A = A\sigma 1 = A$, and $J^2 = -1,0,+1$. 
} 

Quantum mechanics corresponds to $J^2 = -1$ (elliptic composability), classical mechanics corresponds to $J^2 = 0$ (parabolic composability), and the unphysical hyperbolic quantum mechanics corresponds to $J^2 = +1$ (hyperbolic composability).

The composability two-product algebra is a consequence of the algebraic constraints due to the invariance of the laws of nature under tensor composition. We can reconstruct the usual quantum mechanics formulation from it in a particular case using deformation quantization. Quantum mechanics can be formulated over real numbers, complex numbers, quaternions, or $SL(2, \mathbb{C})$. In the case of real numbers, complex numbers, or quaternions, quantum mechanics predicts probabilities by using Born rule. In the less known case of $SL(2, \mathbb{C})$ quantum mechanics predictions are 4-current probability densities. This formulation leads to spinors and Dirac's equation. For the sake of simplicity we will only consider the usual formulation of quantum mechanics over complex numbers. In the classical mechanics case we also restrict ourselves to symplectic manifolds. 

In the classical mechanics case the the product $\sigma$ is commutative and associative. Hence it is isomorphic with regular function multiplication $f\sigma g = fg$. From Darboux theorem \cite{PoissonBook} we obtain that the product $\alpha$ is a bracket: 

\begin{equation}
f \alpha g = \{ f , g \} = f \overleftrightarrow{\nabla} g = \sum_{i = 1}^{n}  \frac{\partial f}{\partial q^i} \frac{\partial g}{\partial p_i} - \frac{\partial f}{\partial p_i} \frac{\partial g}{\partial q^i} .
\end{equation}

From classical mechanics we can now use deformation quantization to construct the Hilbert space formulation. However it is intructive to see how the this is achieved in flat $\RR^{2n}$ space.

First we can build a K\"ahler manifold as follows. In elliptic composability we have a parameter $J$ satisfying $J^2 = -1$. Because the Hamiltonian formalism is defined over the real numbers, $J$ cannot be a scalar and must have a matrix representation. 

To simplify the problem we consider a one-dimensional physical system. In this case the maximum matrix dimension for the representation of $J$ is two. The only possibility is for $J$ to have the same representation as the representation of the complex numbers imaginary unit: 

\begin{equation}
i = J = 
\begin{pmatrix} 
0 & -1 \\ 1 & 0 
\end{pmatrix} .
\end{equation}

In general, $J$ is not the imaginary complex number unit but a tensor of rank $(1,1)$: $J = J^I_{~J}$ with the following matrix representation:  

\begin{equation}
J^I_{~J} = \begin{bmatrix} 0 & -1_n \\ 1_n & 0 \end{bmatrix} .
\end{equation}

In turn $J$ allows to build a metric tensor and an indefinite inner product. We arive at a special case of a K\"ahler manifold: a complex projective space. 

On $\RR^{2n}$ the Berezin quantization \cite{BerezinQuantization} is the following prescription to construct compact operators from continuous functions on phase space:

\begin{equation}
Q_{\hbar} (f) = \int_{\RR^{2n}} \frac{dp dq}{2 \pi \hbar} f(p,q) |\Phi_{\hbar}^{(p,q)} \rangle \langle \Phi_{\hbar}^{(p,q)} | ,
\end{equation}

\noindent where $\Phi_{\hbar}^{(p,q)}$ are coherent states defined as:

\begin{equation}
\Phi_{\hbar}^{(p,q)} = {(\pi \hbar)}^{-1/4} e^{-ipq/2 \hbar} e^{ipx/\hbar} e^{-{(x-q)}^2 /2 \hbar} .
\end{equation}

At this point we recovered the Hilbert space formulation of complex quantum mechanics. This can be double checked by using the Gelfand-Naimark-Segal (GNS) construction \cite{GNSReference} after extracting the C*-algebra condition for any bounded operators $T$ as follows:

\begin{eqnarray}
 {||T \Phi||}^2 = \langle T \Phi, T \Phi \rangle =  \langle T^* T \Phi , \Phi \rangle \leq \\ \nonumber
||T^* T \Phi ||~||\Phi|| \leq ||T^* T||~||\Phi||^2\\ \nonumber
{||T||}^2 \leq ||T^* T|| \leq ||T^*|| ||T|| = {||T||}^2\\ \nonumber
||T^* T|| = {||T||}^2 .
\end{eqnarray}  

The $\RR^{2n}$ case is special and there are other quantization methods available. The approach can be generalized if we start from Poisson manifolds. Any Poisson manifold admits a deformation quantization \cite{KontsevichPaper} but in this general case we may not obtain a K\"ahler manifold.

Next we ask how to handle the case of quantum systems which do not have a classical analog like the case of spin. Spin quantum systems are finite dimensional and in this case a recent result showed that the composability two-product algebra is enough to uniquely recover the Hilbert space \cite{KapustinA}. The full reconstruction problem of quantum mechanics is still open and in general we can obtain either C*-algebras or C*-Hilbert modules. The complete number system classification for quantum mechanics is another open problem.

In the reconstruction cases obtained so far both composition and information considerations are essential in deriving quantum mechanics. The role of composition arguments is unsurprising given Bell's theorem because for a single particle there are classical models which reproduce exactly all quantum mechanics predictions. Pure composition arguments are unable to eliminate non-physical theories like hyperbolic quantum mechanics.  

Composition arguments are special cases of categorical considerations which we proved that determine the algebraic relationships. Information theoretical considerations demand positivity as a pre-requisite, and positivity can recover the norm properties in Hilbert space formulation in certain cases. 

We have seen that the distinction between quantum and classical mechanics has a composition (algebraic), not an information origin. Why does nature prefer elliptic composition over parabolic composition? This is no different from asking why nature prefers having a maximum speed limit over unlimited speeds. For quantum mechanics the answer cannot be given by either categorical or positivity arguments and we have to appeal to experimental evidence.


\section*{References}
\begin{thebibliography}{99}
\bibitem{EPR1} Einstein~A, Rosen~N and Podolsky~B 1935 Can quantum-mechanical description of physical reality be considered complete? {\it Phys.~Rev.} {\bf 47} (10) 777 
\bibitem{Bell1} Bell~J~S 1964 On the Einstein--Poldolsky--Rosen paradox {\it Phys.} {\bf 1} 195
\bibitem{PR1} Popescu~S and Rohrlich~D 1994 Nonlocality as an axiom {\it  Found.~Phys} {\bf 24} 379
\bibitem{MMExperiment} Michelson~A~A and Morley~E~W 1887 On the relative motion of the Earth and the luminiferous ether {\it Am.~J.~Sc.} {\bf 34} 333
\bibitem{GrginCompPaper} Grgin~E and Petersen~A 1976 Algebraic implications of composability of physical systems {\it Comm.~Math.~Phys.} {\bf 50} (2) 177
\bibitem{LandsmanBook} Landsman~N~P 1998 {\it Mathematical Topics Between Classical and Quantum Mechanics} (New York: Springer)
\bibitem{ConnesBook} Connes~A 1994 {\it Noncommutative Geometry} (San Diego: Acad.~Press)
\bibitem{BerezinQuantization} Berezin~F~A 1975 General concept of quantization 
{\it Commun.~Math.~Phys.} {\bf 40} 153
\bibitem{GrginDualityPaper} Grgin~E and Petersen~A 1974 Duality of observables and generators in classical and quantum mechanics {\it J.~Math.~Phys.} {\bf 15} (6) 764
\bibitem{AlfsenShultz} Alfsen~E~M and Shultz~F~W 2003 {\it Geometry of State Spaces of Operator Algebras} (Boston: Birkh\"{a}user)
\bibitem{LodayAlg} Loday~J--L 1992 {\it Cyclic homology} (Berlin: Springer)
\bibitem{PoissonBook} Laurent-Gengoux~C, Pichereau~A and Vanhaecke~P 2013 {\it Poisson Structures} (Heidelberg: Springer)
\bibitem{SahooPlank} Sahoo~D 2004 Mixing quantum and classical mechanics and uniqueness of Planck's constant {\it J.~Phys.}~A {\bf 37} (3) 997
\bibitem{AdlerQuaternions} Adler~S~L 1995 {\it Quaternionic Quantum Mechanics and Quantum Fields} (New York: Oxford Univ. Press)
\bibitem{HyperbolicQMPaper} Khrennikov~A and Segre~G 2006 Von Neumann uniqueness theorem doesn't hold in hyperbolic quantum mechanics {\it Int.~J.~Theor.~Phys.} {\bf 45} (10) 1869
\bibitem{AspectExperiment} Aspect~A, Grangier~P and Roger~G 1982 Experimental realization of Einstein--Podolsky--Rosen--Bohm gedankenexperiment: a new violation of Bell's inequalities {\it Phys.~Rev.~Lett.} {\bf 49} (2) 91
\bibitem{GNSReference} Arveson~W 1981 {\it An Invitation to C*--Algebra} (New York: Springer-Verlag)
\bibitem{KontsevichPaper} Kontsevich, M. 2003 Deformation quantization of Poisson manifolds. {\it Lett.\ Math.\ Phys.} \textbf{66} (3), 157
\bibitem{KapustinA} Kapustin, A. 2013 Is quantum mechanics exact?
{\it J.\ Math.\ Phys.} \textbf{54}, 062,107
\end{thebibliography}
\end{document}